\documentclass{aa}
\usepackage{graphicx}
\usepackage{txfonts}
\usepackage{natbib}
\begin{document}
\title{A new measurement of the cosmic X-ray background}
\author{
A. Moretti\inst{1}, C. Pagani\inst{2}, G. Cusumano\inst{3}, S. Campana\inst{1},  M. Perri\inst{4}, A. Abbey\inst{5}, M. Ajello\inst{6,7}, A.P. Beardmore\inst{5}, D. Burrows\inst{2}, G. Chincarini\inst{1,8}, O. Godet\inst{5}, C. Guidorzi\inst{1,8}, J.E. Hill\inst{9,10}, J. Kennea\inst{2}, J. Nousek\inst{2}, J.P. Osborne\inst{5}, G. Tagliaferri\inst{1}} 

\offprints{alberto.moretti@brera.inaf.it}

\institute{
INAF, Osservatorio Astronomico di Brera, Via E. Bianchi 46, I-23807, Merate (LC), Italy
\and
Department of Astronomy \& Astrophysics, Pennsylvania State University, 525 Davey Lab, University Park, PA 16802, USA
\and
INAF, Istituto di Astrofisica Spaziale e Fisica Cosmica Sezione di Palermo, Via U.\ La Malfa 153, I-90146 Palermo, Italy
\and
ASI Science Data Center, via G.\ Galilei, I-00044 Frascati, Italy
\and  
University of Leicester, LE1 7RH, UK
\and
Stanford Linear Accelerator Center, 2575 Sand Hill Road, Menlo Park, CA 94025
\and
KIPAC, 2575 Sand Hill Road, Menlo Park, CA 94025
\and 
Universit\`a degli Studi di Milano-Bicocca, Dipartimento di Fisica, Piazza delle Scienze 3, I-20126 Milano, Italy
\and 
NASA/Goddard Space Flight Center, Greenbelt Road, Greenbelt, MD20771, USA
\and 
Universities Space Research Association, 10211 Wincopin Circle, Suite 500, Columbia, MD, 21044-3432, USA
%
}
\date{Received ; accepted }
\date{Received ; accepted }
\titlerunning{Measurement of the Cosmic X-ray Background}
\authorrunning{Moretti et al.}  
\abstract{} 
{We present a new measurement of the cosmic X-ray background (CXRB) in
  the 1.5-7 keV energy band, performed by exploiting the Swift X-ray
  telescope (XRT) data archive. We also present a CXRB spectral model in a
  wider energy band (1.5-200 keV), obtained by combining these data with the
  recently published Swift-BAT measurement.}
{From the XRT archive we collect a complete sample of 126 high
  Galactic latitude gamma-ray burst (GRB) follow-up observations. This
  provides a total exposure of 7.5 Ms and a sky-coverage of $\sim$7
  square degrees which represents a serendipitous survey, well suited
  for a direct measurement of the CXRB in the 1.5-10 keV interval.
  Our work is based on a complete characterization of the instrumental
  background and an accurate measurement of the stray-light
  contamination and vignetting calibration.}
{We find that the CXRB spectrum in the 1.5-7 keV energy band can be
  equally well fitted by a single power-law with photon index
  $\Gamma$=1.47$\pm$0.07 or a single power-law with photon index
  $\Gamma$=1.41$\pm$0.06 and an exponential roll-off at 41 keV.  The
  measured flux in the 2-10 keV energy band is 2.18$\pm$0.13
  $\times$10$^{-11}$ erg cm$^{-2}$s$^{-1}$deg$^{-2}$ in the 2-10 keV
  band.  Combining Swift-XRT with Swift-BAT (15-200 keV) we find that,
  in the 1.5-200 keV band, the CXRB spectrum can be well described by
  two smoothly-joined power laws with the energy break at 29.0$\pm$0.5
  keV corresponding to a $\nu F_{\nu}$ peak located at 22.4$\pm$0.4
  keV.}
{Taking advantage of both the Swift high energy instruments (XRT and
  BAT), we produce an analytical description of the CXRB spectrum over
  a wide (1.5-200 keV) energy band.  This model is marginally
  consistent with the HEAO1 measurement ($\sim$10\% higher) at energies
  higher than 20 keV, while it is significantly (30\%) higher at low
  energies (2-10 keV).}
\maketitle
\section{Introduction}
The Cosmic X-ray background (CXRB) is usually defined as the
integrated emission of all the extragalactic sources in the X-ray
energy band ($\sim$2-100 keV). The name background comes directly from
the first X-ray astronomical observation \citep{Giacconi62}, when an
apparently diffuse background was observed together with the first
extra-solar X-ray source (Sco X-1).  The CXRB spectral properties,
flux and isotropy were accurately (10\%) measured over a wide energy
band by the A2 and A4 experiments on board the High Energy
Astronomical Observatory 1 (HEAO1) satellite.  The analytical model
produced by \cite{Gruber99}, combining A2 and A4 observations with
higher energy data has been considered as a reference for many years
(G99 model hereafter).
Much effort has been spent to quantify the fraction of CXRB emission
due to unresolved point sources.  As predicted by \cite{Giacconi87} a
combination of large and deep surveys performed by focusing telescopes
in the soft part of the X-ray spectrum ($<$ 10) keV has succeeded in
resolving almost the entire (80-90\%) CXRB, the resolved fraction
decreasing at higher energies \citep{Moretti03,Worsley05}.  The point
sources producing the resolved fraction of CXRB in the 2-10 keV band
have been found to be mostly AGN with a small contributions from
galaxy clusters and starburst galaxies
\citep{Horn00,Bauer04,Brandt05,Tozzi06}.  Furthermore, a highly
anisotropic diffuse component is present at energies lower than 1
keV \citep{Soltan07}.  This is contributed by the Local Bubble, the
Galactic halo \citep{Galeazzi07} and the intergalactic medium
\citep{Cen99}, while at higher energies, and high Galactic latitude
the diffuse component is negligible.

There is a general consensus on the sources from which the CXRB
originates, and the background paradox can be considered solved
\citep{Setti89}; nevertheless, the spectrum of the X-ray integrated
emission is still very important in the study of the statistical
properties of those sources that are too faint to be detected
individually by currently operating telescopes, as highly absorbed
AGNs and very high red shift quasars.  The extrapolation of the AGN
observed spectra (unabsorbed and Compton-thin) to the region of the
CXRB peak ($\sim$30 keV) can account for $\sim$ 75\% of the peak
\citep{Gilli07}.  Compton-thick AGNs are thought to be responsible for
the remaining fraction \citep{Treister05,Gilli07,Ballantyne07}.  Given
that even the most recent AGN surveys in the hard band, $>$10 keV, can
add only few percent to this number
\citep{Sazonov07,Sazonov08,Tueller08}, the CXRB provides a key
boundary condition in the determination of the census of the heavily
obscured AGNs.
Moreover, an accurate measurement of the CXRB spectrum is an important
observational constraint in the study of the very high redshift
(z$>$6) AGNs which will remain unresolved even with the next
generation of X-ray telescopes \citep{Salvaterra07,Rhook08}.  Finally,
a proper characterization of the CXRB spectrum is also crucial to
ensure proper background subtraction in the study of low surface
brightness diffuse X-ray emission coming from the outskirts of
clusters and groups of galaxies \citep{Gastaldello07,Snowden08}.

Measurements performed in the soft part of the CXRB spectrum with different
instruments (see Table~\ref{tab:cxrb_soft}) yield a scatter which is much 
larger than the one expected from standard candle flux measurements \citep{Kirsch05}, 
meaning that the observed discrepancy cannot be entirely explained by
the differences in absolute calibrations of the individual instruments.
The large scatter and the poor control on systematic uncertainties in
the CXRB measurements led some authors \citep{Ueda03,Treister05} to use
the G99 model, re-normalized by a factor of $\sim$30\%. The underlying
(but not verified) assumption is that the G99 spectrum is correct in
shape but with the normalization affected by some calibration problems
of the HEAO1 instruments. \citet{Worsley05,Worsley06} use an even more
artificial solution, combining the XMM-Newton CXRB measurement
\citep{Deluca04} up to 8 keV and the re-normalized G99 at higher
energies.
On the other hand, recently published measurements, performed by means
of wide-field not-focused hard X-ray instruments (SAX-PDS,
INTEGRAL-IBIS, Swift-BAT) yield results consistent (10\% level) with the G99
model in the 20-50 keV range \citep{Churazov07, Frontera07, Ajello08},
reversing the recent trend that prefers higher intensities \citep{Ueda07}.
\begin{table}
\tabcolsep=2mm
\begin{center}
\caption[]{A compilation of CXRB flux measurements in
  the soft energy band sorted by year of publication, compared with
  the G99 model.}
\begin{tabular}{|l|c|l|}
\hline
Instrument     &Flux 2-10 keV                     &  Reference \\  
               &\scriptsize{[10$^{-11}$erg cm$^{-2}$s$^{-1}$deg$^{-2}$]}  &            \\ 
\hline
HEAO1          & 1.65$\pm$0.17      & \tiny{\citet{Gruber99}    }           \\
Rockets        & 2.20$\pm$0.20      & \tiny{\citet{Mccammon83}  }          \\
ASCA-SIS       & 1.92$\pm$0.09      & \tiny{\citet{Gendreau95}  }         \\ 
SAX-MECS       & 2.35$\pm$0.10      & \tiny{\citet{Vecchi99}    }           \\
ASCA-GIS       & 1.94$\pm$0.20      & \tiny{\citet{Kushino02}   }          \\
RXTE-PCA       & 1.64$\pm$0.05$^*$  & \tiny{\citet{Revnivtsev03}}       \\
XMM-Newton     & 2.24$\pm$0.16      & \tiny{\citet{Deluca04}    }           \\
HEAO1-A2       & 1.66$\pm$0.08$^*$  & \tiny{\citet{Revnivtsev05}}       \\
Chandra        & 2.19$\pm$0.26      & \tiny{\citet{Hickox06}    }           \\
\hline
\end{tabular}
\label{tab:cxrb_soft}

{\noindent (*)The original values are 1.91$\pm$0.06 and 1.96$\pm$0.10 respectively. We correct them, according to
the values reported in  Table 3 of \citep{Revnivtsev05} to account for the cross-calibration
with XMM-Newton.}
\end{center}
\end{table}

Here we present a new measurement of the CXRB spectrum in the 1.5-7
keV energy band, obtained by the analysis of the archival data of the
X-ray telescope (XRT) on board the Swift satellite \citep{Gehrels04},
a mission dedicated to the study of gamma-ray bursts (GRBs) and their
afterglows.  XRT uses a Wolter I mirror set, originally designed for
the JET-X telescope \citep{Citterio94}, to focus X-rays (0.2-10 keV) onto a
XMM-Newton/EPIC MOS CCD detector \citep{Burrows05}.  GRBs are detected
and localized by the Burst Alert Telescope (BAT)\citep{Barthelmy05},
in 15-300 keV energy band and followed-up at X-ray energies (0.3-10 keV)
by the X-Ray Telescope.  Following-up gamma-ray burst afterglows, the
Swift-XRT obtains deep exposures on random positions of the sky,
totally uncorrelated with already known bright X-ray sources,
providing us with a simple and direct measurement of the CXRB
spectrum.
\section{Work strategy}
\begin{figure}
\includegraphics[width=\columnwidth]{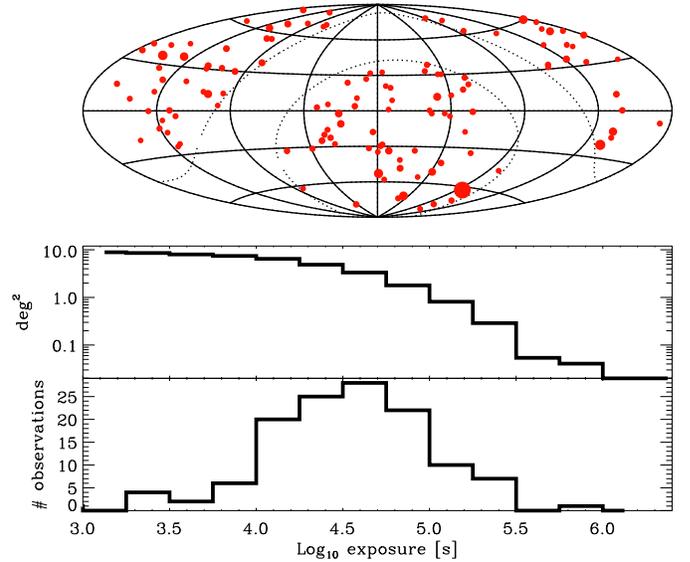}
\caption{
{\bf Upper panel}: Sky distribution of the 126 observations in Equatorial coordinate,
with the Galactic plane exclusion (dotted lines). The size of the points is 
proportional to the exposure and it is not representative of the observed field
size.
{\bf Middle panel}: the cumulative survey sky coverage as a function of (logarithm of) exposure time. 
{\bf Lower panel}: distribution of the (logarithm of) nominal exposure time. 
}
\label{fig:f1}
\end{figure}

For each energy E we can consider the signal S$_{\rm tot}$ registered in a
typical high Galactic latitude GRB afterglow follow-up observation, as
the sum of 4 factors.  These are the GRB signal, the CXRB itself,
which is the one we aim to measure, the electronic and particle
induced background (NXB or instrumental background) and the
stray-light (SL), i.e. the contamination from sources outside the
telescope field of view.
\begin{equation}{\rm
S_{tot}(E)=CXRB(E)\times f_{vign}(E)+ NXB(E)+SL(E)+GRB(E).
  }\label{eq:strategy1}
\end{equation}
The GRB afterglow signal can be easily eliminated by filtering data
both in space and in time.  The CXRB itself is contributed by bright
resolved plus faint unresolved sources and it is affected by
vignetting (f${\rm_{vign}}$).
We measure the NXB using two different and independent datasets: a two-
day observation performed with the focal plane camera assembly
(FPCA) sun-shutter closed (SC) and the data collected in a region of
the detector which is not exposed to the sky (NES).
We evaluate the third element, the SL contamination
using a series of off-axis observations of bright sources.
Given the high level of  the CXRB isotropy \citep{Shafer83,Revnivtsev08} 
we can consider this factor as a fraction of the CXRB (SL =
f$_{\rm{SL}}\times$CXRB) in such a way that:
\begin{equation}{\rm
CXRB(E)= \frac{S_{tot}(E) - NXB(E)} {f_{vign}(E)+ f_{SL}(E)}.
}\label{eq:strategy2}
\end{equation}
To measure the CXRB spectrum we perform stacked spectral analysis of a
large sample made only of GRB follow-up observations.  
The following four sections are devoted to fully describing the
technical details of our work, which allowed us to estimate the terms
of Eq.~\ref{eq:strategy2} and their uncertainties.  In particular, in
Section~\ref{sect:sample} we describe the dataset and the reduction
procedures; in Section~\ref{sect:NXB} and ~\ref{sect:stray} we present
how we measure the NXB and SL contamination, respectively.  In
Section~\ref{sect:vigne} we describe how we calculate the vignetting
correction and its uncertainty.  In Section~\ref{sect:spec} we present
the spectral analysis procedure and its results, discussing them in
Section~\ref{sect:disc}.

Throughout this paper, all errors are quoted at 90\% confidence level,
unless otherwise specified. The photon index is denoted as $\Gamma$.
\section{Sample selection and data reduction}
\label{sect:sample} 
\begin{figure}
\includegraphics[width=8.0cm]{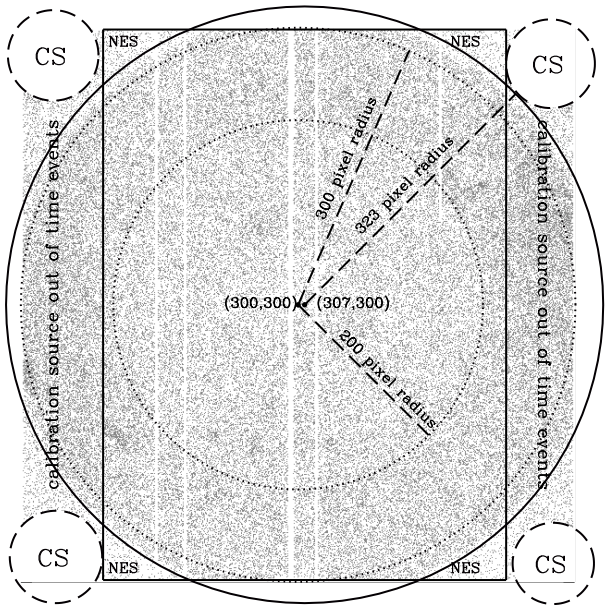}                                          
\caption{XRT detector. The inner dotted circle indicates the
detector region we consider for the CXRB measurement. The larger dotted
circle indicates the nominal field of view. The continuous circle shows
the conservative definition of field of view we use to define the NES regions.
The continuous box is used to define the NES border and show the regions 
contaminated by calibration source (CS) out of time events.}
\label{fig:f2}
\end{figure}
During the first 45 months of operation (Jan 2005- July 2008), the Swift-XRT
observed some 300 GRB afterglows with typical exposure times of
70-100 ks during the $\sim$ 10 days following the prompt event.  We
consider all the long (T$_{90}>$ 2s) GRB follow-up observations with
a nominal standard exposure longer than 10 ks and Galactic latitude
$|b| > 20^{\circ}$.  Because we find long term
variations of the NXB level we consider only data after January 2006
in such a way that the NXB scatter remains lower than 10\% (see
Section~\ref{sect:NXB}).  Similar variations ($\lesssim$3\% per year)
in the NXB level were observed in SAX-LECS-MECS and ASCA-GIS
\citep{Perri00,Kushino02} and were interpreted as due to a gradual
drop in the satellite altitude and/or to the cycle of solar
activity.

The sample consists of 126 GRB observations from January 2006 to July
2008.  For each observation we exclude from our analysis the data
collected in the first day (the segment 0) in order to exclude the
brightest phases of the afterglows.  For each observation we consider
only the central 200 pixel radius (7.9 arcminutes) circle, excluding a
30 pixel radius (1.18 arcminutes), corresponding to $\sim$95\% of the
encircled energy fraction of a point-like source \citep{Moretti05} 
around the GRB
position.  This corresponds to a nominal field of view (FOV) of 0.054
square degrees.  The real observed sky solid angle varies from
observation to observation depending on the precise pointing distribution
of the observation.

We reduce data using the standard software (HEADAS
software, v6.4, CALDB version Dec07) and following the procedures
in the instrument user
guide \footnote{http://heasarc.nasa.gov/docs/swift/analysis/
  \#documentation}.  We replace the standard good time interval
(GTI) definition, which is tuned for the observations of bright point-like sources,
by more restrictive filters. Due
to the failure of the thermo-electric cooler power supply, the XRT CCD temperature
is subject both to orbital (4$^\circ$C in 5.9 ks) and long term
(15$^\circ$C on a day time scale) variations, ranging between -70$^\circ$C and -47$^\circ$C
\citep{Kennea05}.  Dark current and hot pixels are highly temperature
dependent and create high instrumental background in the low energy
band (0.3-0.7 keV) during observations performed at temperatures
higher than -52$^\circ$C \citep{Pagani07,Moretti07}.  Moreover, due to the low
orbit of the Swift satellite, a typical target can be observed no more than 1-2
ks on a single orbit. Therefore the data from single object are split in
different segments.  Occasionally some reflected light from the Earth
limb significantly increases the very low energy ($< 0.5$~keV) background 
at the end or at the beginning of an observation
segment. To reduce these effects, we select intervals with CCD temperature $<$-55$^\circ$C
and elevation angle (i.e. the altitude of the observation direction on
the Earth horizon) $>$ 40$^\circ$, instead of the standard
-47$^\circ$C and 30$^\circ$, respectively.  Moreover, we consider only
data from observation segments longer than 300 seconds and eliminate
the first and the last 100 seconds of each orbital segment.
After the complete time-filtering procedure, these procedures typically reduce 
the effective exposure time to 50\% of the standard ones.  
The total nominal exposure time considered is 7.5 Ms with median value of 40 ks for
single observations.  
The final exposure time distribution of the
126 observations, together with the sky-coverage is shown in
Fig.~\ref{fig:f1}: the surveyed area is 7 and 1.3 deg$^2$ 
at 10 and 100 ks respectively.
\section{Instrumental and particle induced signal}
\label{sect:NXB}
\begin{figure}
\includegraphics[width=9.0cm]{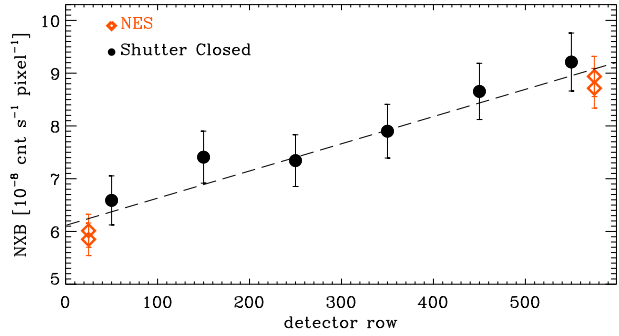}                                          
\caption{NXB count rate (per pixel) in the 1.5-7 keV energy band as
  measured by SC and NES datasets. The linear gradient in the CCD
  vertical direction is evident.}
\label{fig:f3}
\end{figure}
To evaluate the instrumental and particle induced background (NXB) we 
use two different datasets.

First, we use the two day observation performed between 2007-09-04
18:50:00 UT and 2007-09-06 18:42:00 UT when the instrument Sun shutter
(0.38 mm thickness stainless steel of grade 302) was closed due to an
improper slew which brought the XRT to point $\sim$ 15 degrees from
the Sun.  The instrument automatically closed the shutter in front of
the camera. For the next two days the usual XRT observations were
performed, but with the shutter closed.  We apply to these data the
same reduction and filtering procedures that we apply to the sky data.
The final exposure time for the instrumental background with the
shutter closed is 67 ks with an average count rate of
1.92$\pm$0.05$\times$10$^{-7}$ (7.71$\pm$0.3$\times$10$^{-8}$) counts
s$^{-1}$pixel$^{-1}$ in the 0.3-10 (1.5-7.0) keV energy band.  For the
remainder of the paper we will refer to this dataset as shutter closed
(SC) dataset.
\begin{figure}
\includegraphics[width=9.0cm]{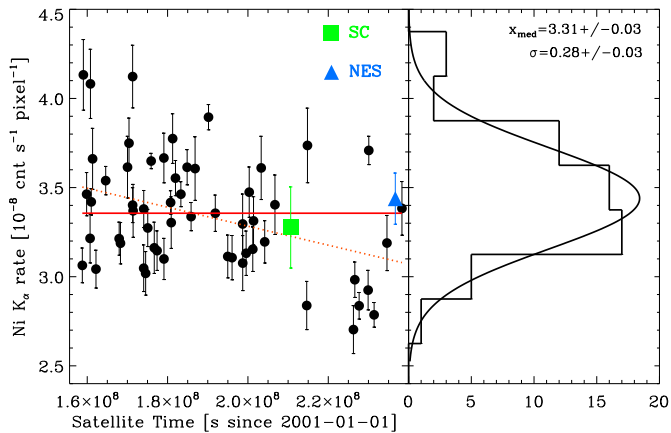}                                          
\caption{The instrumental Ni K$_{\alpha}$ line flux registered in the
  56 observations with exposure times longer than 40ks. The continuous
  line is the average, while the dotted line is the best linear
  fit. The total scatter of these measurements with respect of the
  average is at the level of 10\%(1 $\sigma$), as shown in the right
  panel of the figure.}
\label{fig:f7}
\end{figure}

The second dataset is provided by the data collected in the regions of
the detector which are not exposed to the sky (NES). These are four
different regions (2507 pixels each) close to the CCD boundary and
delimited by the FOV and the corner sources (Fig.~\ref{fig:f2}).  The
FOV and corner source region definitions are reported in the standard
calibration file (CALDB) \texttt{swxregion20010101v003.fits}.  In
particular the nominal field of view of the telescope is the 300 pixel
radius circle centered on the detector pixel (300,300).  We
conservatively adopted a wider definition of the FOV which is the 323
pixel radius circle centered in the detector pixel (307,300). Then, we
define the NES as the parts of the box centered in detector pixel
(307,300), width 436 and height 596 lying outside the conservative
FOV\footnote{In simpler words, according to ds9 syntax, this
  corresponds to \texttt{box(307,300,436,596,0)-circle(307,300,323)}}.
The signal registered in these regions has been telemetered since June
2008 when the telemetered detector area was increased to
600$\times$600 pixels.  We use all the available data present in the
Swift-XRT archive of the photon counting (PC) observations between
June-July 2008.  This results in a total exposure of 2.4 Ms.
\begin{figure}
\includegraphics[width=9.0cm]{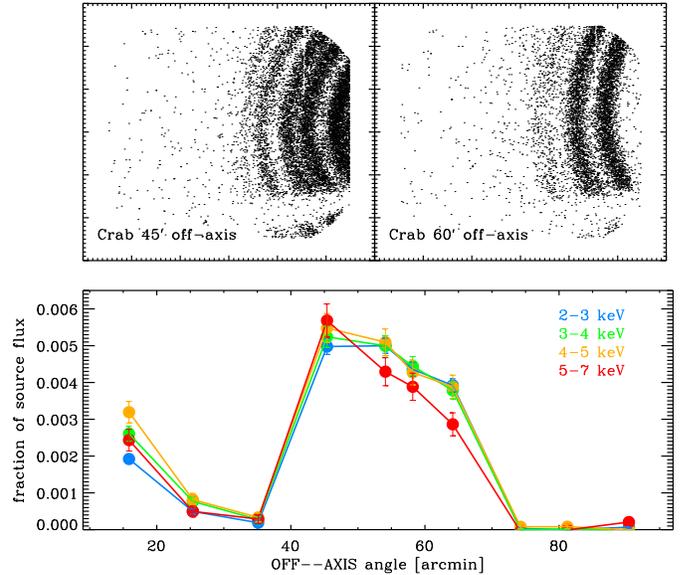}                                          
\caption{{\bf Upper Panel:} off-axis observations of the Crab nebula, used to
calibrate the SL contamination. {\bf Lower Panel:} ratio between
SL and on-axis flux from the Crab at different off-axis angles.}
\label{fig:f4}
\end{figure}
The uncertainty in the determination of NXB is one of the main sources
of uncertainty of this measurement.  For both these datasets we assume
that the signal is contributed only by the particle induced and the
pure instrumental background.  Possible sources of systematic errors
are the following.  First, the fluorescence from the shutter itself:
if the shutter produces some fluorescence lines, our NXB
estimate would be systematically higher than the correct value.
Second, significant inhomogeneities in the CCD response or in the
intrinsic fluorescence background, could bias the NXB measurement in
the NES regions.  The third error source is the time dependence of the
NXB. As already mentioned, in some previous missions, where the NXB
background has been estimated by means of observation of the dark
Earth, a time dependence on time scale of years has been observed.

To check our data for these systematic errors, we first verify
consistency between the two datasets, SC and NES. Comparing the two
datasets we find that the NXB in the 1.5-7 keV band displays a
gradient in the vertical direction of the CCD, with the bottom regions
being 30$\%$ lower than the top regions.  This trend is very well
described by a linear fit with SC and NES datasets being highly
consistent (Fig.~\ref{fig:f3}).  The consistency of the two datasets
give us good confidence that the first two sources of systematics are
negligible for our purposes.  We do not have a direct way to monitor
the NXB time dependence during all the observations (NES data started
to be telemetered only in June 2008) over the entire energy
band. Nevertheless we can use the data in the 7-8 keV interval, where
the CXRB signal is low and the detected signal is dominated by the
Ni K$_\alpha$line at 7.48 keV produced by the fluorescence of the
telescope material.  This is uniformly distributed over the detector
area with a typical count rate of 3.8$\times$10$^{-8}$ count
s$^{-1}$pixel$^{-1}$.  We compare the Ni line observed in the NES
regions with the one observed during the SC and sky observations.  To
minimize the statistical error we consider the 56 observations longer
than 40ks.  We model the data in the 6.8-8.2 keV energy band by means
of a Gaussian plus a straight line. We find that the Ni line flux
decreases slightly with mission time (correlation at 2.5$\sigma$
confidence) producing scatter of $\sim$ 10$\%$ with respect to the
average (Fig. \ref{fig:f7}).  We note that the line flux registered in
SC and NES data is perfectly consistent with the average of the
observed fluxes.  For the purpose of the stacked spectral analysis we
account for this uncertainty in the final error budget as explained
below in Sect.~\ref{sect:spec}.
\section{Stray-light}
\label{sect:stray}
\begin{figure}
\includegraphics[width=\columnwidth]{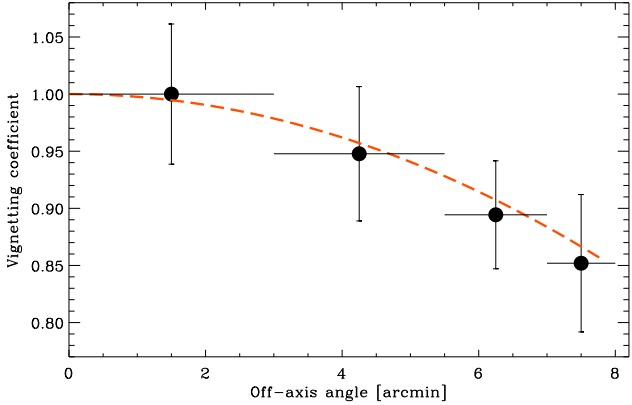}
\caption{Variations of the effective area as function of the off-axis angle, usually
called vignetting. The dashed line represents the standard calibration function, while
the black points are the values we find for the 2-3 keV energy band.}
\label{fig:fv}
\end{figure}
The other main component of the non-cosmic background is the SL.
Ray-tracing simulations indicate that this is produced by
photons coming from sources that are outside the telescope FOV at
distance between 20 and 100 arcminutes from the optical axis of the
telescope, whereas the FOV of the telescope (mirror + detector) has a
radius of $\sim$ 12 arcminutes.  A fraction of the photons produced by
these sources reach the detector after only one reflection on the
mirror or even directly, passing through the mirror shells without any
interaction.  Some X-ray telescopes mount baffles on top of the
mirrors to prevent such a contamination .  This is not the case for
XRT, for which the SL is a significant fraction of the diffuse
radiation registered on the CCD.

In order to evaluate the level of contamination in XRT images, we take
advantage of the many Crab Nebula calibration observations.  Then a
series of observations at 7 different off-axis angles, ranging from 15
to 90 arcminutes were performed.  We compare them with the on-axis
calibration observation. We calculate the fraction of the source flux
present on the central 200 pixel radius circle of the detector as
the ratio between the flux observed at each distance from the optical
axis and the on-axis flux.  Given the large off-axis angles the
dimensions of the Crab nebula are negligible for our purposes.  The
results of this analysis are shown in Fig.~\ref{fig:f4}. We split our
analysis in several different energy bands, finding no significant
variation as function of energy up to 5 keV.  These observations
clearly show that XRT images are contaminated by the emission of
sources outside the field of view up to $\sim$ 70 arcminutes.  Due to
the isotropy of the CXRB we can calculate the expected contamination
as the surface integral of the (measured) relative flux produced by
the Crab observed at different off-axis angles.  The result of the
integration in the 1.5-7 keV band is the CXRB fraction ${\rm f_{SL}}$=
0.268$\pm$0.015 (see Eq.~\ref{eq:strategy2}).

To give an idea of the SL contamination in absolute terms, using the
LogN-LogS calculated by \cite{Moretti03}, we find that for each XRT
image we expect a contaminating flux of 2.7$\pm 0.1 \times$
10$^{-13}$erg s$^{-1}$ diffuse over the 200 pixel central circle in
the 1.5-7 keV band, corresponding to a count rate of 6.3$\pm 0.2
\times$10$^{-3}$ (assuming a spectral photon index of 1.4).
\section{Vignetting}
\label{sect:vigne}
Because the CXRB is an extended source with a
uniform surface brightness profile, the variation of the effective
area as function of the off-axis angle, i.e. the vignetting, must be
accounted for.  For each energy the vignetting correction can be
analytically described by a polynomial function \citep{Tagliaferri04}.
Therefore, first, we calculate the vignetting correction as function
of the off-axis angle using the standard calibration (CALDB coefficients
\texttt{swxvign20010101v001.fits}).  Then, the total vignetting
factor ($f_{{\rm vign}}$(E) in Eq.~\ref{eq:strategy1}) is given by the
integration of this function over the 200 pixels radius circular
region. In the 1.5-7 keV energy band the integrated vignetting
correction ranges from 6\% to 14\%.

\begin{figure}
\includegraphics[width=\columnwidth]{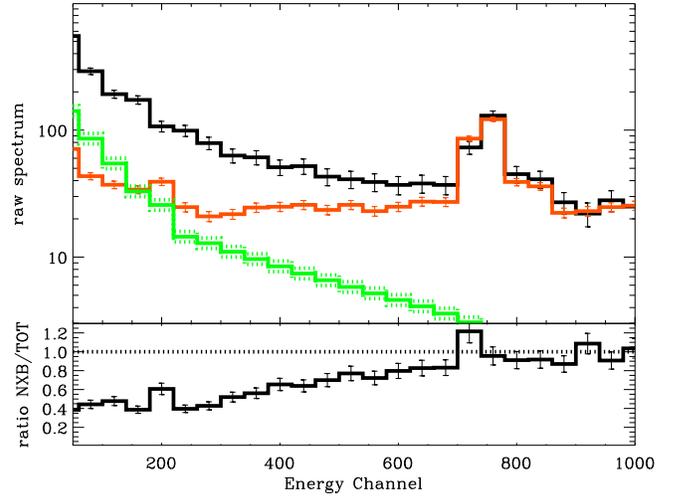}                           
\caption{The different components of the signal registered during an
  observation with no bright source present in the field of view.  The
  black line is the whole signal (S$_{\rm{tot}}$) registered during
  the 35 ks observation of the afterglow of GRB080319C.  Data have
  been reduced and filtered as explained in Section~\ref{sect:sample}
  The red line is the NXB component (NES data).  The green line is the
  expected SL contamination. In the lower panel the ratio between the
  total NXB (instrumental+particle induced + SL) and the
  S$_{\rm{tot}}$ is plotted.  The XRT energy channel are 0.1 keV wide.}
\label{fig:f5}
\end{figure}
To estimate the vignetting calibration accuracy, which is a relevant
factor in the uncertainty calculation of our measurement, we consider
all the point-like sources (1945 sources) detected in our survey,
excluding GRB afterglows. We calculate the median count rate of these
sources in different off-axis angle bins. Assuming that the source
populations detected at different off-axis angles coincide
(once we eliminate GRB afterglows) and having the necessary statistics
to make cosmic variance negligible, the ratio between the median
count-rate on- and off-axis give us the vignetting coefficient.  We
repeat this operation for different energy bands and we find that the
standard vignetting calibration is accurate at the level of a few
percent (Fig.~\ref{fig:fv}).
\section{Spectral analysis}
\label{sect:spec}
The XRT nominal energy band is from 0.3 to 10 keV.  The fraction of
S$_{\rm {tot}}$ due to non-cosmic background (NXB+SL) during a typical
XRT observation (without any bright source in the FOV) depends on
energy and is shown in Fig.~\ref{fig:f5}.  In the energy range between
1 and 6 keV the NXB contributes $\sim$50\% of the total signal
registered during the observation.  As a comparison, in the XMM-Newton
data the cosmic fraction of the total signal is 20\% \citep{Deluca04}.
Above 7 keV the NXB and, in particular, the particle induced
background becomes dominant, due to the presence of the Ni
(K$_{\alpha}$ and K$_{\beta}$ at 7.478 and 8.265 KeV) and Au
(L$_{\alpha}$ at 9.713 keV) fluorescence lines. We only consider the
energy band between 1.5 and 7 keV, excluding the data with energy less
than 1.5 keV because the Galaxy contribution is not negligible
\citep{Kushino02, Hickox06}.

We account for the vignetting by modifying the nominal ancillary
response file (ARF,\texttt{swxpc0to12s0(6)20010101v010.arf}) by the
$f_{{\rm vign}}$(E) factor (see Section~\ref{sect:vigne}) .  We modify the ARF
file also to account for the SL contamination according to
Eq. ~\ref{eq:strategy2} with the results reported in
Section~\ref{sect:stray}.  Finally, we calculate the contamination of
the GRB afterglow residuals outside the 30 pixel radius using the
analytical PSF model \cite{Moretti05}.  The GRB residual signal is
energy dependent and contributes a maximum of 1.1\% below 2 keV,
becoming negligible above 3 keV.  We account for this contamination
applying another small energy dependent correction to the ARF file.
To summarize, for each energy E we modify the nominal ARF according:
\begin{equation}{\rm 
ARF'(E)= ARF(E)\times\left( f_{vign}(E) + f_{SL}(E) \right)\times (1-GRB_{res}(E)).
}\label{eq:arf}
\end{equation}
\begin{figure}
\includegraphics[width=\columnwidth]{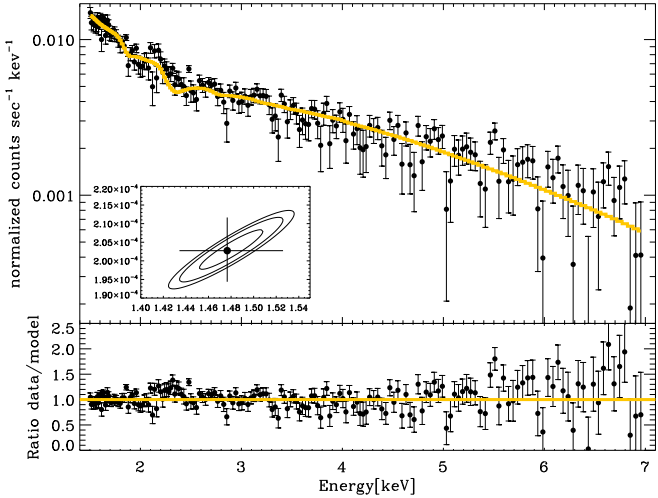}
\caption{CXRB spectrum and the power law best fit. Best fit values are
  reported in Table~\ref{tab:totres}}
\label{fig:f6}
\end{figure}
In order to account for the CCD defects and the excluded 30 pixel
radius circular region around the GRB position we create an exposure
map for each observation. The overall correction, weighted for the
exposure time of the single observation, correspond to 6.3$\%$ for the
sky exposure and 3.2$\%$ for the background.  We correct this
by modifying the BACKSCAL keyword in the spectrum
(\texttt{PHA}) files.  For the background file we consider the NES
regions that provide better statistics than the SC observations
(Section ~\ref{sect:NXB}).  The four NES regions are not homogeneous
due to the spatial gradient in the NXB. Nevertheless the symmetry of
the geometry allows us to use this dataset without any correction.  We
perform the stacked spectral analysis merging the 126 event files,
re-sorting the events and the GTIs, and extracting the spectrum from
the 200 pixel radius central circle in detector coordinates.  To
fit the data and calculate the fluxes we use \texttt{XSPEC}(v12.4).
\begin{figure}
\includegraphics[width=\columnwidth]{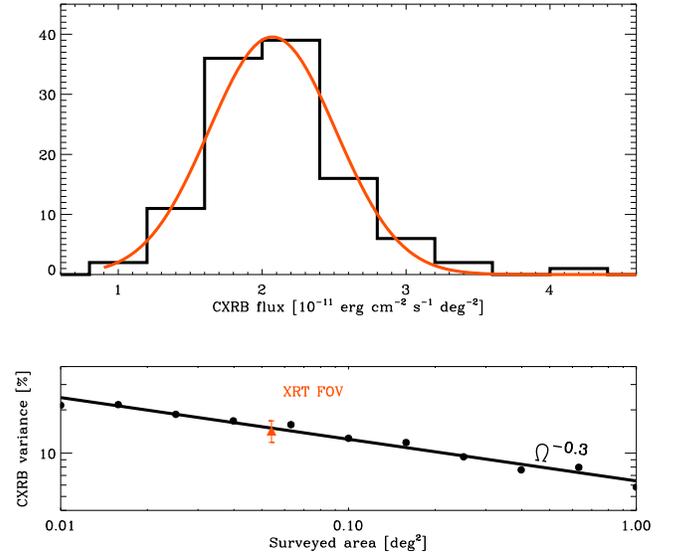}
\caption{{\bf Upper panel:} the distribution of the CXRB flux
  measurements from the 113 observations with effective durations longer than 10
  ks (black line), together with the best Gaussian fit (red line).
  The scatter in the measurements is the sum of the statistical error
  ($\sim$15\%) and cosmic variance.  {\bf Lower panel:} The comparison
  between CXRB variance expected (continuous line and small circles)
  with the observed one, corrected for the statistical contribution
  (triangle).}
\label{fig:f8}
\end{figure}

To account for the systematic uncertainties in the NXB (Section
~\ref{sect:NXB}), vignetting factor (${\rm f_{vign}}$, Sect.~\ref{sect:vigne})  
and SL contamination (${\rm f_{SL}}$, Sect.~\ref{sect:stray}) measurement
we produce a large number (10,000) of simulated datasets, randomly
varying the NXB normalization, the ${\rm f_{vign}}$ and ${\rm
  f_{SL}}$, according to the appropriate Gaussian distributions.
For the NXB normalization we use the mean standard deviation that we
observe for the Ni line fluxes in our sample (lower panel of
Fig.~\ref{fig:f4}).  For the ${\rm f_{vign}}$ and ${\rm f_{SL}}$ we
conservatively use a standard deviation equal to 5\% which slightly exceeds
the estimated errors.

We neglect the Galactic contribution (absorption and emission) and fit
our data by means of a simple power law, obtaining the numbers
reported in Table~\ref{tab:totres} and shown in Fig.~\ref{fig:f6}.  For
analogy with previous works in the literature \citep{Gruber99,Frontera07},
we also fit our data with a cut-off power law (CPL) with the energy
break fixed to 41.13 keV.
Both the models provide a good description of our data in the 1.5-7 keV
energy interval. 

\begin{figure}
\includegraphics[width=\columnwidth]{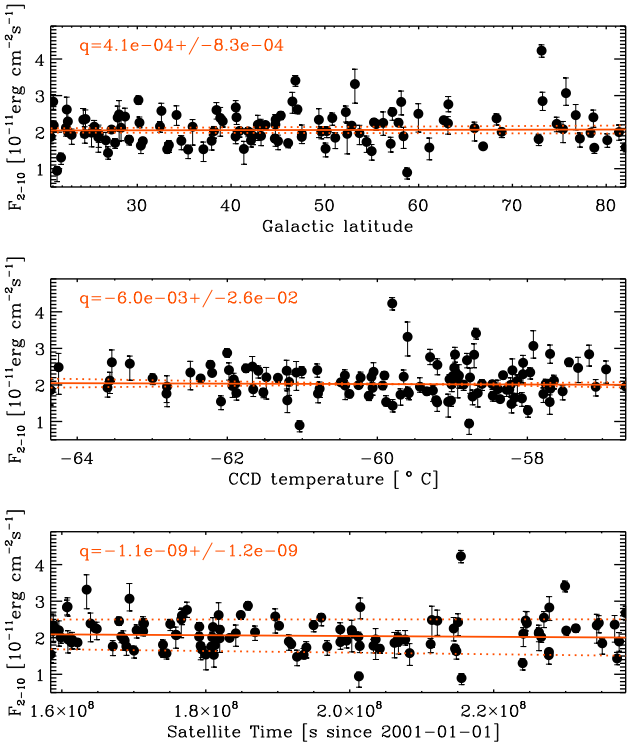}
\caption{CXRB measures plotted against time, Galactic latitude and CCD
  temperature. Here the plotted error are only the statistical ones.
  We fit all the three datasets with a straight line (y=qx+a), finding
  that they are consistent with a constant (q=0) at the level of 1
  $\sigma$.}
\label{fig:f9}
\end{figure}

Finally we combine our data with the Swift-BAT CXRB measurement 
performed in the energy band 20-150 keV \citep{Ajello08}
and we fit the joined energy distributions
with two smoothly joined power laws (2SJPL) of the form
\begin{equation}{\rm
E^2~\cdot \frac{dN}{dE} = \frac{C~\cdot~E^2}{(E/E_B)^{\Gamma_1}
  +(E/E_B)^{\Gamma_2}}~[keV~cm^{-2}~s^{-1}~deg^{-2}].  }\end{equation}
This is the same model \citet{Ajello08} uses to fit a large collection
of CXRB measurements, together with the Swift-BAT new measurement.  We
note that, because a specific response matrix has been produced in the
Swift-BAT measurement, the cross-calibration factor for point-like
sources \citep{Godet09} cannot be applied here.

Our best fit values are reported in Table~\ref{tab:totres}.
The peak of the energy distribution E$_{{\rm peak}}$ is given by 
\begin{equation}
{\rm E_{peak}= E_B \cdot \left( \frac{2-\Gamma_2}{\Gamma_1-2}\right)^{\frac{1}{\Gamma_1-\Gamma_2}} = 22.4\pm0.4~keV}.
\end{equation}
\begin{table*}
\begin{center}
\caption[]{The best fit results for the three different models we
  use. Among brackets are reported the
  statistical contribution to the total error budget.}
\begin{tabular}{|l|c|c|c|c|c|}
\hline
Model&Range [keV]&Norm.\scriptsize{[keV$^{-1}$cm$^{-2}$s$^{-1}$deg$^{-2}$]} &Ph.ind.($\Gamma$)&Flux 2-10 \scriptsize{[erg cm$^{-2}$s$^{-1}$deg$^{-2}$]}&\scriptsize{$\chi^2$}(dof)\\
\hline
& & & & &  \\
PL &1.5-7 &3.69$_{-0.20(0.18)}^{+0.20(0.18)}\times$10$^{-3}$&1.47$_{-0.07(0.06)}^{+0.07(0.06)}$&2.18$_{-0.13(0.02)}^{+0.12(0.02)}\times$10$^{-11}$&209.1(175)\\  
& & & &&  \\
\hline
& & & &&  \\
CPL &1.5-7&3.70$_{-0.20(0.18)}^{+0.20(0.18)}\times$10$^{-3}$&1.41$_{-0.06(0.06)}^{+0.06(0.06)}$&2.13$_{-0.13(0.02)}^{+0.13(0.02)}\times$10$^{-11}$&209.7(175)\\  
& & & &&  \\
\hline
\hline
Model&Range [keV]&Norm.\scriptsize{[keV cm$^{-2}$s$^{-1}$sr$^{-2}$]} &$\Gamma_1$, $\Gamma_2$, ${\rm E}_{\rm B}$ &Flux 2-10\scriptsize{keV[erg cm$^{-2}$s$^{-1}$deg$^{-2}$]}&$\chi^2$(dof)\\
\hline
& & & &&  \\
2SJPL &1.5-200&0.109$_{-0.003}^{+0.003}$keV&1.40$_{-0.02}^{+0.02}$, 2.88$ _{-0.05}^{+0.04}$,29.0$_{-0.5}^{+0.5}$&2.21$_{-0.07}^{+0.07}\times$10$^{-11}$&200.5(193)\\   
&& & &&  \\
\hline
\end{tabular}
\label{tab:totres} 

\end{center}
\end{table*}
\subsection{Cosmic variance}
To study the variance of our sample we consider the 113 observations
with durations longer than 10 ks for which the spectral parameters and
the flux of the CXRB can be calculated with an acceptable accuracy
($\sigma_{\rm s}\sim$15\%).  We find that the flux distribution is well described by
a Gaussian with a standard deviation of $\sigma_{\rm o}$=20.8\%$\pm$2.4 (upper panel of
Fig.~\ref{fig:f8}).  The maximum CXRB flux value in our sample, at
$\sim$ 5 $\sigma_{\rm o}$ from the mean, is observed in the field of GRB
071028B where the galaxy cluster Abell S1136 is present.  This is not
surprising, given the fact that Abell clusters are $\sim$5000
distributed over $\sim$27,000 deg$^2$ of sky meaning that 1-2 Abell
clusters are expected in our survey.

The variance we observe in the flux distribution is contributed by
both by the statistical error ($\sigma_{\rm s}$) and cosmic variance
($\sigma_{\rm c}$).  We find that the latter is consistent with the CXRB
variance expected for the area surveyed by a single observation if we
assume that the CXRB is entirely produced by point sources. In fact,
it can be shown that, if we assume that the source fluxes are
distributed as the classical F$^{-3/2}$ LogN-LogS, the cosmic variance
scales with the surveyed area as $\Omega^{-0.5}$,
\citep{Revnivtsev08}.  Assuming the LogN-LogS calculated by
\cite{Moretti03}, which is flatter at low energies and
generating 1000 random samples with different dimensions (ranging from
0.01 to 1 square degree of sky), in the flux range
10$^{-16}$-10$^{-10}$ erg cm$^{-2}$s$^{-1}$, we find that a more
realistic value is $\Omega^{-0.3}$, as shown in the lower panel of
Fig.~\ref{fig:f8}.  We find that, with 0.054 deg$^2$ XRT field, we
expect a variance of 15.1\%. This is very close to the one we observe
in our sample, once we account also for the contribution of the
statistical error: ${\rm \sigma_c^2=\sigma_o^2-\sigma_s^2}$=14.3$\pm$1.8\%.
We note also that the extrapolation of our
simulations to the total surveyed area ($\sim$7 deg$^2$) tells us that
the stacked analysis uncertainty due to cosmic variance is negligible.
\subsection{Data check}
We check our data against any bias due to the time of the observation,
CCD temperature, and Galactic latitude.  As explained in Section
\ref{sect:NXB} we find a slight dependence of the NXB on the time of
the observations.  However, as the NXB dependence is slight and the
NXB contribution is minor the 113 CXRB flux measurements do not have
any significant correlation with the observation time (see lower panel
of Fig.~\ref{fig:f9}).  As already said (Section ~\ref{sect:sample})
the XRT CCD temperature is variable due to the fact that it is only
passively controlled. Because dark current and hot pixels are
temperature dependent, we also plot the 113 flux measurements against
the average temperature of the observations (see central panel of
Fig.~\ref{fig:f9}).  Finally we check the flux measurements against
Galactic latitude to exclude any significant contribution from our
galaxy to the XRT measurements (upper panel of Fig.  ~\ref{fig:f9}).
For all the three datasets we find that the best linear fit is
consistent with a constant at the level of 1 $\sigma$ .
\section{Discussion}
\label{sect:disc}
\begin{figure*}
\includegraphics[width=16cm]{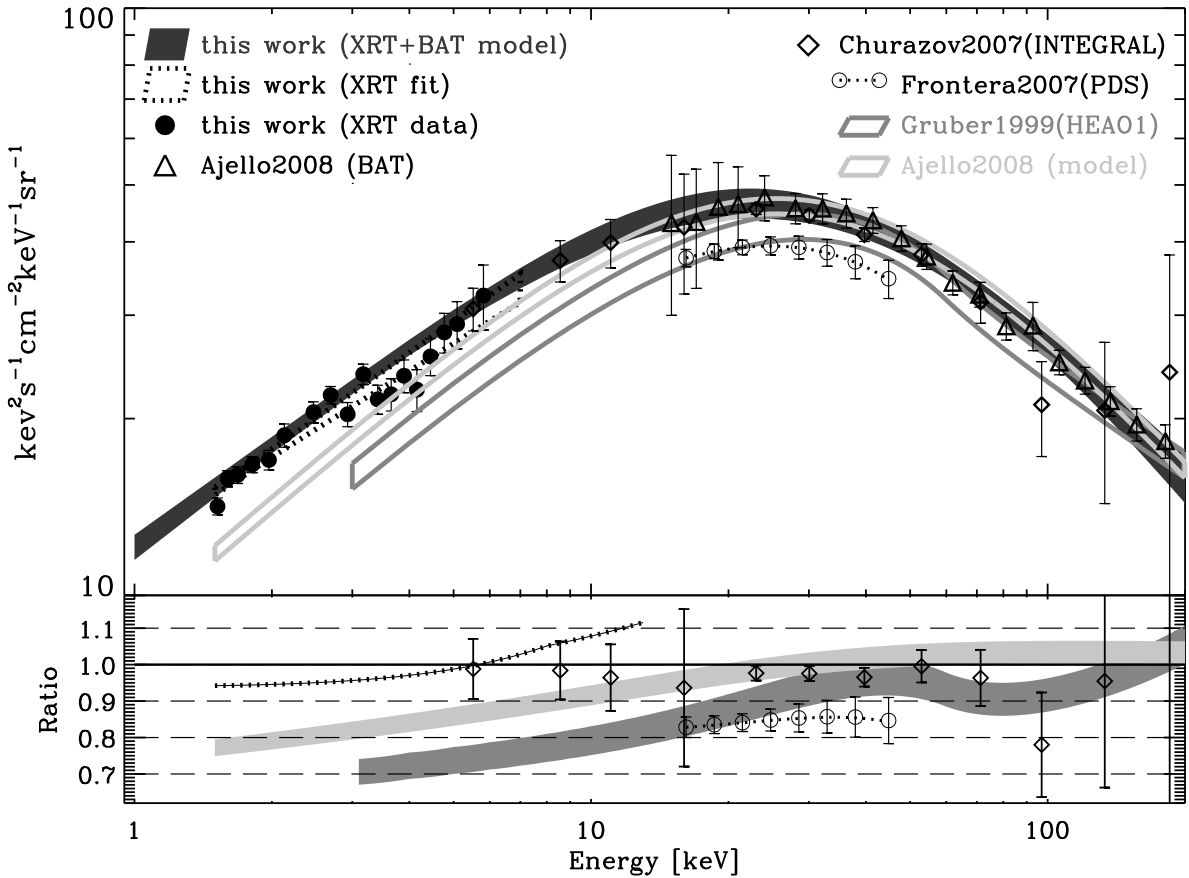}
\caption{{\bf Upper panel:} A compilation of flux measurements both in
  the soft and hard energy bands. For the clearness of the plot, not
  all the soft energy measurements reported in
  Table~\ref{tab:cxrb_soft} are shown here.  Because \citet{Gruber99}
  do not report the uncertainties in the best fit values, we use a
  fiducial 5\% error for G99 model.  {\bf Lower Panel:} Ratio of the
  flux measurements plotted in the upper panel with our joined XRT+BAT
  fit. Colors are the same of upper panel. For comparison, with the
  dotted line we plot (only in the bottom panel) the ad-hoc model
  \citet{Worsley05} used to calculate the resolved fraction.}
\label{fig:f10}
\end{figure*}
As mentioned in the Introduction, the CXRB spectrum normalization is
still a debated issue.

The Swift-XRT measurement, we present here, is very close to
XMM-Newton (Table \ref{tab:cxrb_soft}). This is not unexpected as the
Swift-XRT effective area calibration has been slightly modified to match
XMM-Newton by means of simultaneous observations
\footnote{SWIFT-XRT-CALDB-09-V11 available at
  http://heasarc.gsfc.nasa.gov/docs/heasarc/caldb/swift/docs/xrt/index.html}.
For what concerns cross-calibration, Swift-XRT measures fluxes 5-10\%
lower than RXTE-PCA during simultaneous observations of 3c273 \citep{Godet09}. 
Cross-calibration observations of 1E
0102.2-7219, the brightest supernova remnant in the Small Magellanic
Cloud, recently showed that Chandra-ACIS, XMM-Newton-MOS, Suzaku-XIS
and Swift-XRT agree to within $\pm$10\% for all instruments
\citep{Plucinsky08}.  Therefore, the differences with HEAO1 and
RXTE-PCA measurements cannot be entirely explained by the absoulte
calibration differences, as already pointed out by \citet{Frontera07}.

 In the region of the CXRB peak ($\sim$30 keV) all the measurements show a
acceptable agreement($\sim$10\%), in the soft band the XRT measurement
confirms \citet{Revnivtsev05} conclusions: in the 2-10 keV band
narrow-field focusing telescopes measure CXRB values which are
significantly higher than the ones found by wide-field not focusing
telescopes.  However, the XRT data, although inconsistently higher
than the G99 model, smoothly join the higher energy data as we show by
the good fit to the XRT and BAT.

Below 60 keV, the G99 model consists in a CPL with $\Gamma$=1.29 and
energy break 41.13 (note that no uncertainties are reported in the
\citet{Gruber99} paper).  As shown in Fig.~\ref{fig:f10}, the
differences from the G99 model range from 30\% below 10 keV down to
5-10\% in the region of the CXRB energy peak.  This is due to the fact
that the slope of the soft part in our best fit is steeper (1.41
instead of 1.29) and the peak of the spectrum is much softer (22 keV
instead of 29 keV).  As previously discussed, \citet{Ajello08} uses a
2SJPL to fit Swift-BAT data together with a large collection of
different CXRB measurements down to 2 keV.  In comparison to this
model we find that the soft energy slope is significantly softer
(1.41$\pm$0.02 versus 1.32$\pm$0.02), while the high energy slope, the
energy break and the normalization are consistent. Interestingly, our
model has the same CPL shape, with energy break at 41.13 keV and
photon index 1.4, that provides the best fit to SAX-PDS data
\citep{Frontera07} in the 20-50 keV band, albeit with a significant
difference in normalization.  Finally we also note that XRT data and
our model are well consistent with the INTEGRAL measurement
\citep{Churazov07} all over the considered energy band.

In summary, breaking the paradigm that G99 spectrum has the correct
shape shows that CXRB data can be analytically described by a
2SJPL with the values reported Table~\ref{tab:totres} and a peak in the
energy distribution at 22.9$\pm$0.4~keV.  In the 1.5-50 keV range,
this function is very similar to a CPL with the energy break of 41.13
keV and photon index of 1.4.

We note that the 2-10 keV CXRB flux measurement directly affects the
calculation of the CXRB resolved fraction.  \citet{Moretti03},
combining shallow and deep surveys and integrating the source number
counts, estimate that the resolved fraction of 2-10 keV CXRB is
87$\pm$6\% .  \citet{Worsley05} refined this calculation finding that
the resolved fraction ranges from 80\% in the 2-4 keV band to
$\lesssim$ 60\% for energies higher than 6 keV.  The main reason for
the inconsistency between the two results is the value of the CXRB
total flux.  \citet{Moretti03} used an average of a sample of CXRB
measurements, yielding a value of 2.02$\pm$0.11$\times$10$^{-11}$ erg
cm$^{-2}$s$^{-1}$deg$^{-2}$ which is 10\% less than the present
measurement.  As already mentioned, \citet{Worsley05} used an {\it ad
  hoc} model, combining the XMM-Newton measurement with a
re-normalized G99 model.  As shown in the Fig.\ref{fig:f10} (bottom panel)
this model, although not motivated from an observational point of view, is not
very far from our best fit.  If we assume the present measurement for
the CXRB and integrate the LogN-LogS of \citet{Moretti03}, we find a
result for the CXRB resolved fraction which is 79$\pm$6\% in the 2-10
keV band, in very good agreement with the average value quoted by
\citet{Worsley05}.  The values relative to the single narrow bands at
higher energies, on the other hand, should be slightly corrected,
applying our CXRB value.

The LogN-LogS extrapolation at very low fluxes (10$^{-17}$ erg
cm$^{-2}$s$^{-1}$deg$^{-2}$, a factor 20 lower than the faintest
Chandra deep field sources) cannot account for all the CXRB. This
implies that a not negligible fraction of the CXRB is supposed to be
produced by non detected sources.  \citet{Worsley06} and
\citet{Hickox07} correlate almost the entire CXRB unresolved fraction
to optical/IR detected galaxies in the Chandra deep fields.  These are
star-forming galaxies which are expected to overwhelm the number of
AGNs at very low fluxes \citep{Ranalli03,Bauer04}, absorbed AGN
\citep{Treister05, Gilli07} which are supposed to be the main
component at higher energies and with a small contribution from very
high redshift (z$>$6) quasars.
\section{Conclusion}
We use the Swift-XRT archival dataset to determine the flux and
spectrum of the CXRB. This has two main advantages.  The first one is
the observational strategy which provides us with a truly random
sampling of the X-ray sky, not correlated with previously known
sources. The second is the low level of the NXB background, which
allows measurement of the CXRB with high accuracy.  Similar to other
focusing telescopes, we find that CXRB flux is significantly higher
than HEAO1/G99 model.  Nevertheless combining our dataset with
Swift-BAT data, we show that we can describe the CXRB spectrum with a
simple model (two smoothly joined power laws) over a wide energy
band. The model we propose is much more observationally motivated than
the ones recently used in the literature for population synthesis
models and for the CXRB resolved fraction calculation.  Using the
present CXRB measurement we calculate that the resolved fraction in
the 2-10 keV energy band is 79$\pm$6\%.
\begin{acknowledgements}
This work is supported at OAB-INAF by ASI grant I/011/07/0, 
at PSU by NASA contract NAS5-00136; AA, AB, OG and JO acknowledge STFC funding. 
This research has made use of NASA's Astrophysics Data System Service
\end{acknowledgements}
\bibliographystyle{aa}   
\bibliography{moretti}   
\end{document}